\documentclass[aps,pra,twocolumn,amsfonts,amssymb,a4paper,showpacs]{revtex4}

\usepackage{graphicx}

\newcommand{\be}{\begin{equation}}
\newcommand{\ee}{\end{equation}}

\newcommand{\ket}[1]{|{#1}\rangle}
\newcommand{\braket}[2]{\langle{#1}|{#2}\rangle}

\newcommand{\nbar}{{\bar n}}

\newcommand{\sumlim}{\sum\limits}

\newcommand{\ehoch}[1]{e^{#1}}

\newcommand{\skipc}[2]{}

\begin{document}

\title{Wave Packets can Factorize Numbers}

\author{Holger~Mack}
\author{Marc~Bienert}
\author{Florian~Haug}
\author{Matthias Freyberger}
\author{Wolfgang P.~Schleich}

\affiliation{Abteilung f\"ur Quantenphysik, Universit\"at Ulm, 89069 Ulm, Germany}


\begin{abstract}
We draw attention to various aspects of number theory emerging in the time evolution of elementary quantum systems with quadratic phases. Such model systems can be realized in actual experiments. Our analysis paves the way to a new, promising and effective method to factorize numbers.
\end{abstract}

\maketitle

\section{Introduction}

Interference of waves instead of Newton's light corpuscles --- what a revolutionary idea of Thomas Young pronounced in his famous talk at the Royal Society in 1801 \cite{bib:young}. The  opposite idea --- particles as waves --- proposed by Louis de~Broglie and cast into the appropriate mathematical form by Erwin Schr\"odinger more than 100 years later opened a new realm in atomic physics with more applications today than ever. ``Can a quantum-mechanical description of physical reality be considered complete?'' This famous question raised by Albert Einstein, Boris Podolsky and Nathan Rosen in their seminal paper \cite{bib:einstein} brought out most clearly the importance of entanglement \cite{bib:schroedinger} as the basic ingredient of quantum theory. Interference, wave mechanics and entanglement are the three ideas on which Peter Shor could build his factorization algorithm \cite{bib:shor1} with a polynomial rather than exponential speed. This discovery gave birth to the new field of quantum information \cite{bib:books}. Guided by the same three central ideas we show in the present paper that the time evolution of elementary quantum systems carries intrinsically the tools to factorize numbers \cite{bib:varenna}. 

Stationary states have played an important role in the development of quantum mechanics. For this reason the time evolution of wave packets has for many years taken a back seat even in textbooks on quantum theory. In this context it is interesting to recall that already in 1927 Earle H.~Kennard \cite{bib:kennard} studied the time evolution of wave packets in simple potentials \cite{bib:nieto}. Recently, sophisticated techniques to create short laser pulses have opened a new era for a controlled creation and observation of wave packets in atoms \cite{bib:yeazell}, molecules \cite{bib:vrakking} and cavities \cite{bib:eberly}. Even wave packet dynamics of cold atoms \cite{bib:raithel} and Bose-Einstein condensates \cite{bib:leanhardt} has been observed. These examples illustrate convincingly that today's technology allows to promote this gedankenexperiment of factorization to a real implementation in a quantum system. 

The use of wave packet dynamics to factorize numbers has been recognized for special examples. Reference \cite{bib:dowling} uses a $N$-slit Young interference setup. Hence, the number to be factorized is encoded in the number of slits. The indicator of a factor is the interference pattern on a screen as a function of the wavelength of the light or matter wave. Whenever the intensity across the screen is constant, the appropriately scaled wavelength provides a factor. Similarly, Ref. \cite{bib:harter} considers the time evolution of a particle in a box. In contrast to these contributions we provide a general framework for factorization and moreover, exhaust the potential of interference by capitalizing on properties of quadratic phase factors.

Our paper is organized as follows. In Sec.~\ref{sec:wavepackets} we briefly review the essential ingredients of wave packet dynamics \cite{bib:buch}. In particular, we present an approximation for the autocorrelation function that illustrates its ability to factorize numbers. The richness of the autocorrelation function originates from quadratic phases which are also the origin of the Talbot effect \cite{bib:rohwedder}, revivals and fractional revivals \cite{bib:leichtle1}, curlicues \cite{bib:curlicues} and quantum carpets \cite{bib:physworld}. In Sec.~\ref{sec:quantumrotor} we turn to one of the most elementary quantum systems, namely the quantum rotor. In this case the autocorrelation function is a Gau\ss\ sum which has been discussed extensively in number theory \cite{bib:lang}. We show that under appropriate conditions factors of a number manifest themselves in non-vanishing real or imaginary parts of the autocorrelation function. We explain this phenomenon using the properties of Gau\ss\ sums. Our conclusions and outlook presented in Sec.~\ref{sec:conclusions} summarize the main results and provide an ``idea for an idea'' how to incorporate entanglement into our scheme.

\section{Wave packets}
\label{sec:wavepackets}

In the present section we briefly review the essential features of wave packet dynamics. We start from the definition of the autocorrelation function and approximate the energy spectrum by one that is quadratic in the quantum number. Physical systems such as Rydberg atoms and diatomic molecules are well described by such a spectrum. We then show the possibility of factorizing a number.

\subsection{Autocorrelation function}

In standard wave packet experiments many physical observables of an atom or molecule are accessible. However, one quantity is most appropriate to obtain information about the dynamics and is nowadays measured routinely: It is the so-called autocorrelation function
\be
	S(t)\equiv\braket{\psi(t=0)}{\psi(t)}=\sumlim_{n=0}^\infty|\psi_n|^2\ehoch{-iE_nt/\hbar},
	\label{eq:autocorr}
\ee
which indicates the overlap between the time-evolved wave packet
\be
	\ket{\psi(t)}= \sumlim_{n=0}^\infty\psi_n\ehoch{-iE_nt/\hbar}\ket n
	\label{eq:wavepacket}
\ee 
and the initial state. Here, the coefficients $\psi_n$ denote the probability amplitudes of the initial state in the basis of energy eigenstates $\ket n$ with corresponding energy eigenvalues $E_n$.  

While so far the autocorrelation function has served as a tool for observing the dynamics, our goal is to use its intrinsic features to perform computational tasks. In particular, we show that autocorrelation functions can reveal prime factors of an appropriately encoded number. 

\begin{figure}
\includegraphics{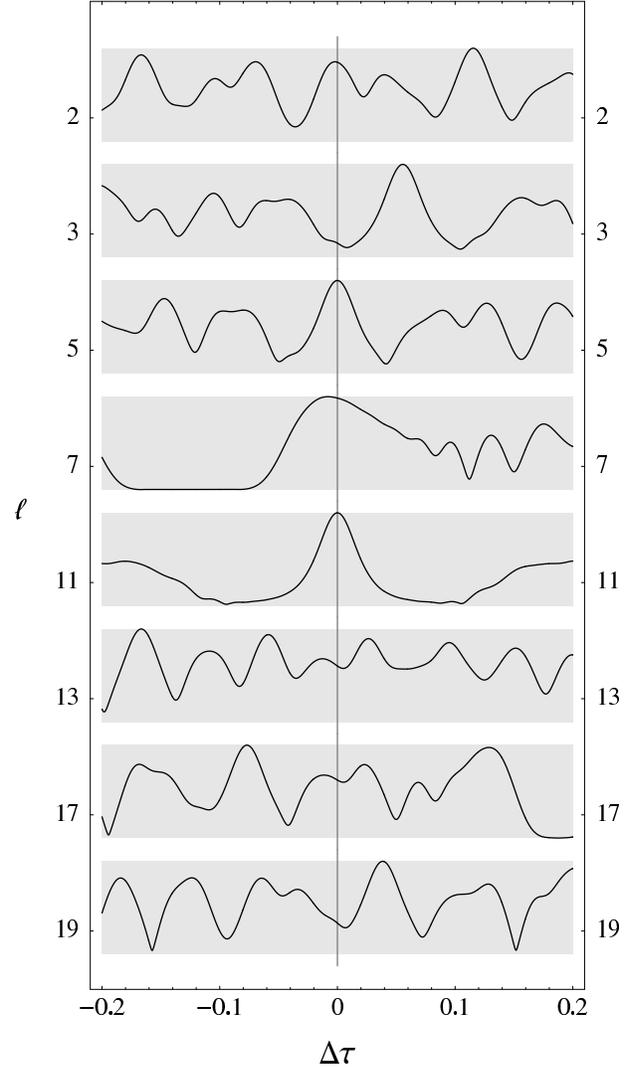}
\caption{Factorization based on the autocorrelation function of a wave packet: a ``guess-the-factors riddle''. We show the absolute value squared $|S_N(\tau=\ell+\Delta\tau)|^2$ of the autocorrelation function, defined in Eq.~(\ref{eq:scaledauto}), as a function of dimensionless time $\Delta\tau$ in the vicinity of various integers $\ell=2,3,5\dots$. The autocorrelation function has a symmetric maximum at an integer, that is at the origin of each horizontal axis, provided this integer is a factor of $N$. In the present case we clearly recognize $5$ and $11$ as factors of $N=55$. The width of the initial wave packet is $\Delta m=10$.
\label{fig:revivals}}
\end{figure}

For this purpose, we consider a wave packet which is localized in energy space around a quantum number $\nbar$, that is, the coefficients $\psi_n$ assume only significant values in a small vicinity of $n=\nbar$. This feature allows us to expand the energy spectrum
\be
	E_n\approx E_{\nbar}+\left.\frac{\partial E_n}{\partial n}\right|_{n=\nbar}(n-\nbar)+\frac12\left.\frac{\partial^2E_n}{\partial n^2}\right|_{n=\nbar}(n-\nbar)^2
	\label{eq:enapprox}
\ee    
around $\nbar$. Here we have assumed that the spectrum is smooth and dense. Moreover, we have omitted higher order terms.

The approximation, Eq.~(\ref{eq:enapprox}), is valid for smooth potentials and a narrow energy distribution of the wave packet. To be specific, we consider the sodium dimer $\textrm{Na}_2$. Here a proper laser pulse lifts the vibrational ground state in the potential of the electronic state $\textrm{X}^{\,1}\Sigma_{\textrm{g}}^+$ to the excited electronic state $\textrm{A}^{\,1}\Sigma_{\textrm{u}}^+$. In the potential curve of the excited electronic state the wave packet is no longer the ground state, but has a small spectral width around a mean value $E_\nbar$. Indeed, in this case third order terms can be neglected \cite{bib:leichtle1}.

To study the wave packet dynamics, Eq.~(\ref{eq:wavepacket}), we use the approximation, Eq.~(\ref{eq:enapprox}), and obtain the phase
\be
	E_nt/\hbar\approx E_{\nbar}t/\hbar+\frac{2\pi t}{T_{\rm cl}}(n-\nbar)+\frac{2\pi t}{T}(n-\nbar)^2.
\ee
Here we have introduced the classical period $T_{\rm cl}$ and revival time $T$.

Thus the autocorrelation function, Eq.~(\ref{eq:autocorr}), reads
\be
	S(t)\approx\ehoch{-iE_{\nbar}t/\hbar}\sumlim_{m=-\infty}^\infty\!\!W_m\exp\left[-2\pi i\left(\frac{m}{T_{\rm cl}}+\frac{m^2}{T}\right)t\right]
\ee
with a shifted summation index $m\equiv n-\nbar$ and the weight function $W_m\equiv|\psi_{m+\nbar}|^2$. Moreover, we have extended the lower bound of the sum to $-\infty$, since in this domain the weight factors $W_m$ do not contribute significantly.

\subsection{An allusion to factorization}

In order to demonstrate the ability of the autocorrelation function to factorize a number, we start from a Gaussian weight function $W_m\propto\exp\left[-m^2/2\Delta m^2\right]$ and fix the ratio $N\equiv T/T_{\rm cl}$ to an integer. In this case, we have to deal with the sum
\be
	S_N(\tau)\equiv\sumlim_{m=-\infty}^{\infty}\!\!W_m\exp\left[-2\pi 			i\left(m+\frac{m^2}{N}\right)\tau\right]
	\label{eq:scaledauto}
\ee
for dimensionless time $\tau\equiv t/T_{\rm cl}$. 

The sum Eq.~(\ref{eq:scaledauto}) is already the key to factorization, although it might not be obvious at this moment. In order to bring this latent capability to light, we focus in Fig.~\ref{fig:revivals} on the modulus squared $\left|S_N(\tau)\right|^2$ of the autocorrelation function for $N=55$. We show this quantity for times $\tau$ in the vicinity of prime numbers $\ell$. Only if $\ell$ is a factor of $N$ the function displays a clear symmetric maximum.

The phenomenon of fractional revivals \cite{bib:leichtle1} lies at the heart of this feature. It is well-known that wave packets can show revivals at multiples of the time $T$ and fractional revivals at times $t=\frac{p}{q} T$ with $p$ and $q$ being mutually prime integers. In our case, this fact implies that at integer multiples of $T_{\rm cl}$ --- or in other words for integers $\tau$ --- we find a fractional revival of the wave packet, that is, a symmetric sequence of Gaussian shaped humps. A detailed analysis of this structure \cite{bib:varenna} brings to light that in most cases neighboring Gaussians overlap and interfere. For this reason, at most times no distinguished revivals appear. Only for times $\tau$ equal to a factor of $N$ the peaks are sharp enough so that the central peak does not overlap with its neighbors and reveals a clear maximum.

\section{Quantum rotor}
\label{sec:quantumrotor}

The crucial point in our factorization scheme is the interference of quadratic phase factors. In the previous section, they were the consequence of an anharmonic potential. But quadratic phase factors even appear for a free particle with a continuous energy spectrum. However, in order to encode integers we need a discrete one. Such a spectrum can be achieved, for example, by periodic boundary conditions. A diatomic molecule rotating perpendicular to its molecular axis is one example of such a quantum rotor.

\subsection{Time evolution}

In order to lay the foundations for the factorization scheme presented in this section we first briefly review the essential properties of the one-dimensional quantum rotor. Then we derive the corresponding autocorrelation function.

The time evolution of the quantum rotor is governed by the Hamiltonian 
\be
	\hat H \equiv \frac{\hat J^2}{2 I}
\ee 
with the operator $\hat J$ of the angular momentum and the moment of inertia $I$. Periodic boundary conditions enforce a discrete spectrum of $\hat J$ with eigenstates $\ket k$ and eigenvalues $\hbar k$ with $k\in\mathbb{Z}$. 

The resulting discrete energy spectrum $E_k=\hbar^2 k^2/(2 I)$ gives rise to the time evolution
\be
	\ket{\psi(t)}= \sumlim_{k=-\infty}^\infty\psi_k\exp\left[-i\hbar\frac{k^2}{2I}t\right]\ket k
	\label{eq:wavepacketp}
\ee
of the quantum rotor.

We model an initially well-localized wave packet using the expansion coefficients
\be
	\psi_k \equiv \left\{
	\begin{array}{ll}
	1/\sqrt{N}&\textrm{\quad for \, $0\leq k<N$}\\
	0&\textrm{\quad otherwise}
	\end{array}
	\right.,
\ee 	
which leads to the explicit expression
\be
	S(t)\equiv S_N(t) = \frac 1 N \sumlim_{k=0}^{N-1}\exp\left[-2\pi i k^2\frac{t}{T}\right]
	\label{eq:sumthefirst}
\ee
for the autocorrelation function. Here we have introduced the revival time $T\equiv 4\pi I/\hbar$. 

Again quadratic phases appear. However, the linear term present in the expression Eq.~(\ref{eq:scaledauto}) is absent.

\subsection{Factorization}

\begin{figure*}
\includegraphics{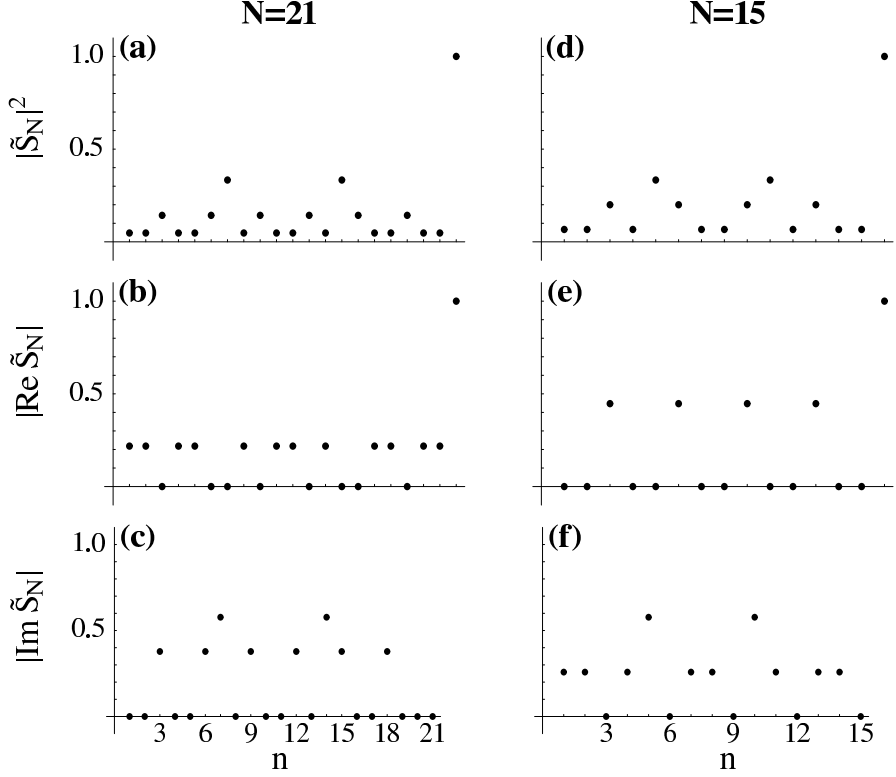}
\caption{Autocorrelation functions of the quantum rotor for $N=21$ (left column) and $N=15$ (right column). From top to bottom, we show modulus squared, real and imaginary part of ${\widetilde S_N}$, Eq.~(\ref{eq:sumthesecond}). For $N=21$ only multiples of both factors $3$ and $7$ show non-zero imaginary parts. For $N=15$ only multiples of the factor $3$ yield a non-zero real part.
\label{fig:gauss}}
\end{figure*}

In order to bring out most clearly the number $N$ to be factorized, we cast the sum, Eq.~(\ref{eq:sumthefirst}), in the form
\be
	{\widetilde S_N}(n) = \frac 1 N \sumlim_{k=0}^{N-1}\exp\left[-2\pi i k^2\frac{n}{N}\right]
	\label{eq:sumthesecond}
\ee
with $n\equiv N\frac{t}{T}\in\mathbb{Z}$. We emphasize that only for integers $n$ the autocorrelation function ${\widetilde S_N}$ is periodic with period $N$.

In Fig.~\ref{fig:gauss}(a) and (d) we show $|{\widetilde S_N}(n)|^2$ for $n=1$ to $N$, with $N=21$ and $N=15$, respectively. We recognize dominant maxima for values of $n$ whenever $n$ and $N$ share a common factor. The appearance of these maxima already indicates that a sum of only quadratic phases allows us to find prime factors of $N$.

We can even go a step further and concentrate on the real and imaginary part of ${\widetilde S_N}$, depicted in Fig.~\ref{fig:gauss}(b) and (c) for $N=21$, and (e) and (f) for $N=15$, respectively. It is remarkable that for $N=21$ only numbers which share a common factor with $N$ appear in the imaginary part, Fig.~\ref{fig:gauss}(c), whereas all the other ones vanish. This is not the case for $N=15$, where only one factor and its multiples appear in the real part, Fig.~\ref{fig:gauss}(e).

\subsection{Gau\ss\ sums}

We recognize that the autocorrelation function, Eq.~(\ref{eq:sumthesecond}), is of the form of a Gau\ss\ sum 
\be
	G(a,b)\equiv\frac1b\sumlim_{m=0}^{b-1}\exp\left[-2\pi i m^2 \frac{a}{b}\right].
	\label{eq:gausstheor}
\ee
Therefore, the structures found in Fig.~\ref{fig:gauss} result from the intrinsic properties of Gau\ss\ sums. In order to highlight this connection, we first briefly review the properties of Gau\ss\ sums and then apply the results to the autocorrelation function Eq.~(\ref{eq:sumthesecond}).

For odd $b$ and mutually prime $a$ and $b$ the sum can be evaluated \cite{bib:lang} and yields
\be
	G(a,b)=\frac{1}{\sqrt{b}}\left(2a \over b\right)\exp\left[i\frac{\pi}{4}(b-1)\right].
	\label{eq:gabjacobi}
\ee
Here we have introduced the Jacobi symbol $\left(\frac{2a}{b}\right)$ which expresses the number-theoretical complexity of the quadratic sum and which can take on the values $\pm 1$.

A closer look at Fig.~\ref{fig:gauss} reveals that the emergence of factors in either the real or imaginary part depends strongly on the representation of the factors as $4s+1$ or $4s+3$ with $s$ being a non-negative integer. Indeed, {\em all} odd numbers $r$ are either of the form $4s+1$ or $4s+3$, and therefore, they are members either of the set
\be
	{\mathcal M}_1\equiv\left\{r\,|\,r=4s+1\right\}
\ee
or
\be
	{\mathcal M}_3\equiv\left\{r\,|\,r=4s+3\right\}.
\ee
 
For the case $b \in {\mathcal M}_1$ the sum $G(a,b)$ is purely real. Indeed, for $b=4s+1$ we find
\be 
	G\propto\exp\left[i\frac{\pi}{4} \,4s\right]=\exp\left[i\pi s\right]=(-1)^s.
\ee 
When $b\in{\mathcal M}_3$, that is $b=4s+3$, we arrive at 
\be
	G\propto\exp\left[i\frac{\pi}{4}(4 s+2)\right]=\exp\left[i\pi \left(s+\frac{1}{2}\right)\right]=(-1)^s i
\ee 
which is purely imaginary. 

Moreover, Eq.~(\ref{eq:gabjacobi}) implies that the modulus 
\be
	|G(a,b)|=\frac{1}{\sqrt{b}}
	\label{eq:modulus}
\ee
is independent of the value of $a$, provided that $a$ and $b$ are mutually prime.  

\subsection{An explanation of the patterns} 

With the help of these properties of the Gau\ss\ sum we can now relate the structures in the autocorrelation function 
\be
	{\widetilde S_N}(n)=G(n,N)
\ee
apparent in Fig.~\ref{fig:gauss} to the factors $p$ and $q$ of $N$. Here we assume that $p$ and $q$ are the only factors of the number $N$, that is $N=p\cdot q$. Moreover, we stipulate that $p$ and $q$ are both odd. For the case of even $N$ one can easily extract factors $2$.

According to Eq.~(\ref{eq:modulus}) the modulus squared $|{\widetilde S_N}|^2$ is given by $1/N$ if $n$ and $N$ do not share a common factor. Otherwise, the common factor cancels in Eq.~(\ref{eq:gausstheor}) and we have $b=1/p$ or $b=1/q$ and thus with the help of Eq.~(\ref{eq:modulus}) we have $|{\widetilde S_N}|^2=1/p$ or $|{\widetilde S_N}|^2=1/q$, respectively. 

We are now in the position to explain the patterns in the real and imaginary parts using the classification of the prime factors into the sets $\mathcal{M}_1$ and $\mathcal{M}_3$. We have to distinguish three cases: (i) $p,q\in{\mathcal M_1}$, (ii)  $p,q\in{\mathcal M_3}$ and (iii) $p\in{\mathcal M_1}, q\in{\mathcal M_3}$ or vice versa. The cases (i) and (ii) imply $N\in{\mathcal M}_1$, whereas (iii) results in $N\in{\mathcal M}_3$.

We start our discussion with an example for case (ii). For $N=21=4\cdot 5+1=3\cdot 7$ both factors $p=3=4\cdot 0+3$ and $q=7=4\cdot 1+3$ are elements of the set $\mathcal{M}_3$ and $N\in\mathcal{M}_1$. If $n$ and $N$ do not share a common factor then $b=N$ and the value of the Gau\ss\ sum is real. Otherwise, we can extract a factor and get $b=q$ or $b=p$, respectively. Consequently, the value of the sum is purely imaginary, since $p,q\in\mathcal{M}_3$. Since the Jacobi symbol mentioned in Eq.~(\ref{eq:gabjacobi}) can only take on $\pm 1$ the corresponding value of $a=n/p$ or $a=n/q$ does not change this behavior. 

The same arguments hold true for the example $N=15=4\cdot 3+3\in\mathcal{M}_3$ illustrating case (iii). Indeed, the factors $p=3=4\cdot 0+3\in\mathcal{M}_3$ and $q=5=4\cdot 1+1\in\mathcal{M}_1$ are not members of the same set. Therefore, if $n$ is a multiple of $p$ we have $b=5$ and thus the sum is purely real. In all other cases, we have either $b=N$ or $b=p$ which are both members of $\mathcal{M}_3$ and yield purely imaginary results. 

The last case (i), not shown in Fig.~\ref{fig:gauss}, discusses factors $p$, $q$ and $N\in\mathcal{M}_1$, resulting in purely real values of the sum for {\em all} $n$. Unfortunately, in this case we do not have vanishing values for integers that are not factors of $N$.

\section{Conclusions and outlook}
\label{sec:conclusions}

The time evolution of a wave packet brings to light key aspects of number theory. This feature allows us to find prime factors $p$ and $q$ of an appropriately encoded number $N=p\cdot q$. These properties are intrinsic to the autocorrelation function, which can be easily measured. At the heart of this phenomenon lie the mathematical properties of Gau\ss\ sums. 

Wave packets of the quantum rotor exhibit these number-theoretical properties in the cleanest form. The most remarkable result is, that in the case of $p,q\in{\mathcal M}_3$ we find all factors (and its multiples) of $N=p \cdot q$ in the imaginary part of the autocorrelation function. All other integers yield a vanishing imaginary part.

But how to make use of these unique features? Imagine that $\textrm{Im}({\widetilde S_N})$ represents the statistics of an observable $\hat O$. Thus, each {\em single} measurement of $\hat O$ would provide one factor of $N$ in a non-probabilistic way. Similarly, in the mixed case $p\in {\mathcal M}_1$ and $q\in {\mathcal M}_3$ the measurement of the real part of ${\widetilde S_N}$ would provide one factor with high probability.

These insights pave the way to an effective method of factorizing numbers. The next step is to find a composite system which (i) intrinsically includes quadratic phase factors (ii) allows parallel calculation of all possible factors and (iii) enables us to find a suitable observable, which reveals all the information about the factors in a deterministic way by use of entanglement.

\acknowledgments

We thank I.Sh.~Averbukh, M.V.~Berry, H.~Maier, I.~Marzoli and F.~Straub, for many fruitful discussions. Moreover, one of us ({W.P.S.}) is most grateful to M.S.~Brandt for organizing a most stimulating Heraeus seminar. The work of {H.M.}, {M.F.} and {W.P.S.} was supported by the Deutsche Forschungsgemeinschaft, by the European Commission through the IST network QUBITS and by the Landesstiftung Baden-W\"urttemberg in the framework of the Quantum Information Highway A8.

\newcommand{\atque}{, and }
\newcommand{\BY}[1]{#1,}
\newcommand{\REVIEW}[4]{{#1} \textbf{#2}, {#3} ({#4})}
\newcommand{\BOOK}[4]{\textit{#1} (#2, #3, #4)}

\end{document}